\documentstyle{mn2e}
\begin{document}

\newcommand\kms{\mbox{${\rm km\,s}^{-1}$}}
\newcommand\cc{\mbox{$\,\rm cm^{-3}$}}
\newcommand\ccs{\mbox{$\,\rm cm^{-3} s$}}
\newcommand\nH{\mbox{$n_{\rm H}$}}
\newcommand\scdm{\mbox{$N_{\rm M}/\Delta V$}}
\newcommand\Td{\mbox{$T_{\rm d}$}}
\newcommand\Tk{\mbox{$T_{\rm k}$}}
\newcommand\X{\mbox{$X$}}
\newcommand\Tb{\mbox{$T_{\rm b}$}}
\newcommand\Nm{\mbox{$N_{\rm M}$}}
\newcommand\beam{\mbox{$\epsilon^{-1}$}}
\newcommand\Wd{\mbox{$W_{\rm d}$}}
\newcommand\WHII{\mbox{$W_{\rm HII}$}}
\newcommand\taud{\mbox{$\tau_{\rm d}$}}
\newcommand\DV{\mbox{$\Delta V$}}
\newcommand\etal{et~al.}
\newcommand\HII{H\,{\sc ii}}
\newcommand\methanol{$\rm{CH_{3}OH}$}
\newcommand\vt{\mbox{$v_{\rm t}$}}
\newcommand\Te{\mbox{$T_{\rm e}$}}
\newcommand\fe{\mbox{$f_{\rm e}$}}

\input epsf
\epsfverbosetrue

\title[Methanol masers at 23.1~GHz]{Search for class~II methanol masers at 23.1~GHz}

\author[Cragg \etal\/]{D.M. Cragg$^1$, A.M. Sobolev$^2$, J.L. Caswell$^3$, S.P. Ellingsen$^4$, and P.D. Godfrey$^1$\\
$^1$ School of Chemistry, Building 23, Monash University, Victoria 3800, 
     Australia;\\
     Dinah.Cragg@sci.monash.edu.au, Peter.Godfrey@sci.monash.edu.au\\
$^2$ Astronomical Observatory, Ural State University, Lenin Street 51, 
     Ekaterinburg 620083, Russia;\\
     Andrej.Sobolev@usu.ru,\\
$^3$ Australia Telescope National Facility, CSIRO, PO Box 76, Epping, NSW 2121, 
     Australia;\\
     jcaswell@atnf.csiro.au\\
$^4$ School of Mathematics and Physics, University of Tasmania, 
     Private Bag 21, Hobart, Tasmania 7001, Australia;\\  
     Simon.Ellingsen@utas.edu.au\\}

\maketitle


\begin{abstract}

In the early days of methanol maser discoveries the $9_{2}-10_{1}\ \rm{A}^{+}$ transition at 23.1~GHz was found to exhibit maser characteristics in the northern star-forming region W3(OH), and probable maser emission in two other sources.  Attention subsequently turned to the 6.6-GHz $5_{1}-6_{0}\ \rm{A}^{+}$ methanol maser transition, which has proved a valuable tracer of early high-mass star formation.  We have undertaken a new search for 23.1-GHz methanol masers in 50 southern star formation regions using the Parkes radiotelescope.  The target sources all exhibit class~II methanol maser emission at 6.6~GHz, with 20 sources also displaying maser features in the 107.0-GHz $3_{1}-4_{0}\ \rm{A}^{+}$ methanol line.  Strong emission at 23.1~GHz in NGC~6334F was confirmed, but no emission was detected in the remaining sources.  Thus the 23.1-GHz methanol masers are rare.

A maser model in which methanol molecules are pumped to the second torsionally excited state by radiation from warm dust can account for class~II maser activity in all the transitions in which it is observed.  According to this model the 23.1-GHz maser is favoured by conditions representing low gas temperature, high external dust temperature, low gas density, and high column density of methanol;  the scarcity of this maser indicates that such combinations of conditions are uncommon.  We have undertaken new model calculations to examine the range of parameters compatible with the upper limits on 23.1-GHz emission from our survey.  Further constraints apply in sources with upper limits to maser emission at 107.0~GHz, and the combination of data for the two transitions delineates a narrow range of gas density and methanol abundance if the dust temperature is 175~K or greater.  While the results are subject to the uncertainties of the chosen model, they may be applicable to the majority of methanol maser sites in the vicinity of newborn high-mass stars, in which methanol masers other than the 6.6- and 12.1-GHz transitions are not detected.

\end{abstract}



\section{INTRODUCTION}

Methanol masers at 6.6 and 12.1~GHz have been detected in several hundred star-forming regions (e.g. Caswell \etal\ 1995a, 1995b), where they are often closely accompanied by OH maser emission and continuum emission from a nearby ultra-compact \HII\ region.  These two exceptionally strong maser lines are the archetypal class~II methanol masers.  Known only in star-forming regions, they probe the early stages in the development of a massive star, which is otherwise shielded from view by the enveloping cloud of dust and gas.  In some sources the strong 6.6-GHz $5_{1}-6_{0}\ \rm{A}^{+}$ and 12.1-GHz $2_{0}-3_{-1}\ \rm{E}$ masers are accompanied by class~II maser emission in up to 20 additional methanol transitions.  The latter are considerably weaker but otherwise display similar characteristics to the strong maser transitions: narrow lines at the same velocity, often displaying multiple components, sometimes variable, with small source size implying high brightness temperature where this has been investigated.  The class~I methanol masers, in contrast, are observed in a different set of methanol lines, and are found offset from other indicators of recent massive star formation.

The $9_{2}-10_{1}\ \rm{A}^{+}$ line at 23.1~GHz was the first methanol maser identified outside the class~I source Orion~KL (Wilson \etal\ 1984), and was the first discovered maser of what later became known as the class~II variety.  The maser nature of the emission towards W3(OH) was confirmed by VLA measurements (Menten \etal\ 1985, 1988a), in which several velocity components were identified as emanating from regions with small angular size, with brightness temperature $>10^7$~K for the strongest feature of intensity 10~Jy and velocity -43.2~\kms.  Emission spectra with similar narrow features in single-dish observations of NGC~7538 at 0.5~Jy (Wilson \etal\ 1984) and NGC~6334F at 52~Jy (Menten \& Batrla 1989) were also identified as probable masers.  Many class~II methanol maser sources were subsequently identified at 6.6~GHz, but no widespread surveys for the 23.1-GHz transition have been published to date, although preliminary results of a VLA survey were presented by Vargas \etal\ (2000).  We report here a new search for 23.1-GHz class~II methanol masers in southern star formation regions.  

There is considerable evidence that class~II methanol masers at different frequencies may be spatially coincident.  VLBI observations of W3(OH) show that the strongest 6.6- and 12.1-GHz features coincide within a maser spot size (0.001~arcsec) and also coincide in velocity (Menten \etal\ 1992).  Furthermore, Menten \etal\ (1988b) found that individual 12.1-GHz maser components in W3(OH) are concentrated towards the most intense 23.1-GHz emission centres (observed with lower resolution).  Such observations suggest that the various maser transitions are excited simultaneously, and so provide constraints on maser pumping models.  To date, only three maser sources have been identified at 23.1~GHz, all of which also exhibit maser emission in the 107.0-GHz $3_{1}-4_{0}\ \rm{A}^{+}$ transition.  The high resolution observations of W3(OH) show correspondence between the emission regions at these two frequencies (Menten \etal\ 1988a, Sutton \etal\ 2001); however, the peak emission at 107.0~GHz stems from the northern edge of the ultra-compact \HII\ region where the corresponding 23.1-GHz maser spot is weak, while the southern edge produces the 23.1-GHz peak together with weaker emission at 107.0~GHz.  This means that the 23.1-GHz line carries new information on the conditions in maser sources, consistent with the two transitions exhibiting different sensitivities to physical conditions in the models.  The 23.1-GHz transition stems from levels which are higher in energy than those involved in most other methanol maser lines, and so weak maser emission is unlikely to be confused by overlapping thermal emission, as can happen for some millimetre methanol maser transitions.

The class~II methanol masers are pumped by infrared radiation from warm dust in the model of Sobolev \& Deguchi (1994).  The pumping proceeds via the \vt=2 and \vt=1 torsionally excited states.  This model successfully accounts for the predominance of the 6.6- and 12.1-GHz masers (Sobolev, Cragg \& Godfrey 1997a), and the existence of many other weaker class~II maser transitions, including the 23.1 and 107.0-GHz lines (Sobolev, Cragg \& Godfrey 1997b).  The weaker maser transitions can be very sensitive to the model conditions, with calculated brightness temperatures changing by many orders of magnitude as the model parameters are varied.  

Many transitions become masers simultaneously in the model.  This provides a probe of the physical conditions in the maser region.  On the one hand it permits best-fitting model conditions to be estimated via multi-transition analysis, when masers at several frequencies are observed in the same source.  Such studies have been undertaken for three sources in which data for many transitions are available (Cragg \etal\ 2001, Sutton \etal\ 2001).  On the other hand, upper limits on nondetected maser lines can also set constraints on the model parameters.  Surveys at different frequencies have detected the weaker maser transitions in only a few of the class~II methanol maser sources; for example, the 107.0-GHz maser is the most common of the weaker maser lines, but is detected in only 25 sources of more than 175 surveyed (Val'tts \etal\ 1995, Val'tts \etal\ 1999, Caswell \etal\ 2000, Minier \& Booth 2002).  Thus the absence of the weaker methanol maser transitions in the majority of class~II methanol maser sources can be used to define the range of conditions under which the 6.6-GHz masers usually develop.


\begin{figure}
\centerline{\epsfxsize=8.5cm\epsfbox{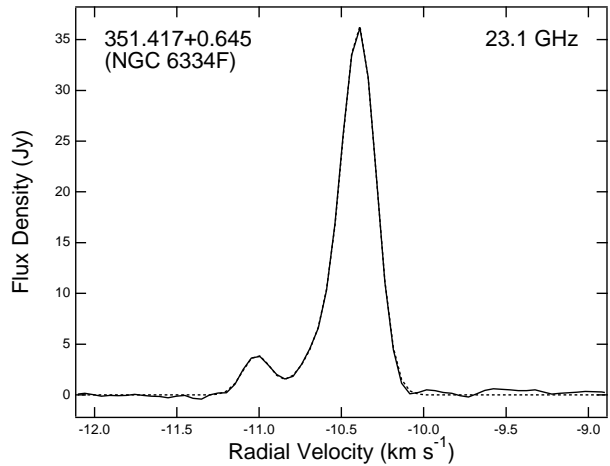}}
\caption{23.1-GHz methanol maser spectrum detected in 351.417+0.645 (NGC~6334F), with three Gaussians fitted (dotted trace).}
\label{fig:ngc_2}
\end{figure}


\begin{table}
\caption{Gaussian components fitted to detected 23.1-GHz maser in 351.417+0.645 (NGC~6334F)}
\label{tab:gfit}

\centerline{
\begin{tabular}{ccc} \hline
     Flux Density  &   Velocity      &     FWHM \\
    (Jy)     &    (\kms)      &    (\kms)  \\
\hline \\
 32.77 &  --10.39  &   0.24\\
  6.35 &  --10.55   &  0.36\\
  3.85 &  --11.01    & 0.20\\
\end{tabular}}

\end{table}


\section{OBSERVATIONS}

Observations were carried out using the Australia Telescope National Facility 64-m Parkes antenna in 2001 October, with an overall system efficiency at 23.1~GHz of 6.0~Jy/K.  The intensity calibration is relative to Virgo A and is estimated to have an error of less than 10 percent. The precise rest frequency adopted for the $9_{2}-10_{1}\ \rm{A}^{+}$ methanol line was 23121.0242~MHz (Mehrotra, Dreizler \& Mader 1985).  The target sources were observed for 5 min at the on-source position, followed by similar reference observations offset by +10 min in RA.  Typically this procedure was repeated until a total on-source integration time of 30~minutes was reached. The correlator configuration of 2048 channels across 8~MHz, in each of 2 circular polarizations, gave a channel spacing at 23.1~GHz of 0.0507~\kms, and velocity coverage 104~\kms. The data were processed using the {\sc spc} reduction package.  After converting to LSR velocities, scans were averaged together, quotient spectra were formed from the appropriate signal and reference, the two polarizations were added, a polynomial baseline fitted and subtracted, and Hanning smoothing applied.  Upper limits to any detected signal were obtained from these spectra, and are pertinent to narrow maser emission where polynomial baselines can safely be removed. In good weather, the typical rms noise of the spectra was 0.3~Jy.

50 southern class~II methanol maser sources were searched at 23.1~GHz. These were selected from the list of sources showing methanol maser emission at 6.6 and (usually) 12.1~GHz (Caswell \etal\ 1995a, 1995b), including the 20 sources with masers at 107.0~GHz  (Caswell \etal\ 2000) which are observable from Parkes.  Narrow maser emission from NGC~6334F previously reported by Menten \& Batrla (1989) was readily detected, and reobserved regularly as a system check.


\begin{table*}

\caption{Upper limits on 23.1~GHz methanol flux density ($3\sigma$) for sources surveyed.  Vel is the centre of the velocity range searched.  For each source the observed peak maser intensity $S$ at 6.6~GHz is tabulated, together with the maser peak or upper limit at 107.0~GHz (Caswell \etal\ 2000).  Where no maser is detected, the upper limit to the brightness temperature ratio $R$ (defined in Section 3) is evaluated for the 23.1 and 107.0~GHz lines relative to the 6.6~GHz peak.} 

\label{tab:sources}

\centerline{
\begin{tabular}{rrrrrrrrr}\hline\\                          
\multicolumn{1}{c}{Source} & \multicolumn{1}{c}{RA}   & \multicolumn{1}{c}{Dec}   & \multicolumn{1}{c}{Vel} & \multicolumn{1}{c}{$S(6.6)$} &  \multicolumn{1}{c}{$S(23.1)$} &  \multicolumn{1}{c}{$S(107.0)$} &  \multicolumn{1}{c}{$R(23.1/6.6)$} &  \multicolumn{1}{c}{$R(107.0/6.6)$} \\
 & \multicolumn{1}{c}{(J2000)}   & \multicolumn{1}{c}{(J2000)}   & \multicolumn{1}{c}{($\kms$)} & \multicolumn{1}{c}{(Jy)} &  \multicolumn{1}{c}{(Jy)} &  \multicolumn{1}{c}{(Jy)} &   &   \\
\hline\\                          
291.274--0.709 & 11 11 53.35 & --61 18 23.8 & --40 & 100 & $<$0.9 & $<$1.4 & $<$7.E--04 & $<$5.E--05 \\
305.200+0.019 & 13 11 16.93 & --62 45 55.1 & --40 & 44 & $<$1.5 & $<$1.4 & $<$3.E--03 & $<$1.E--04 \\
305.202+0.208 & 13 11 10.49 & --62 34 38.7 & --40 & 20 & $<$1.5 & $<$2.4 & $<$6.E--03 & $<$5.E--04 \\
305.208+0.206 & 13 11 13.71 & --62 34 41.4 & --40 & 320 & $<$0.5 & $<$3.8 & $<$1.E--04 & $<$5.E--05 \\
308.918+0.123 & 13 43 01.86 & --62 08 52.1 & --50 & 62 & $<$1.4 & $<$1.4 & $<$2.E--03 & $<$9.E--05 \\
309.921+0.479 & 13 50 41.78 & --61 35 10.1 & --51.4 & 635 & $<$0.7 & $<$1.7 & $<$1.E--04 & $<$1.E--05 \\
310.144+0.760 & 13 51 58.48 & --61 15 42.4 & --51.7 & 130 & $<$0.8 &  23 & $<$5.E--04 &   \\
313.577+0.325 & 14 20 08.59 & --60 42 00.9 & --40 & 150 & $<$1.5 & $<$1.8 & $<$9.E--04 & $<$5.E--05 \\
316.359--0.362 & 14 43 11.20 & --60 17 13.2 & --10 & 96 & $<$1.4 & $<$1.4 & $<$1.E--03 & $<$6.E--05 \\
316.381--0.379 & 14 43 24.20 & --60 17 37.4 & --10 & 38 & $<$1.5 & $<$0.5 & $<$3.E--03 & $<$5.E--05 \\
316.640--0.087 & 14 44 18.44 & --59 55 11.5 & --10 & 128 & $<$1.4 & $<$1.4 & $<$9.E--04 & $<$4.E--05 \\
318.948--0.196 & 15 00 55.40 & --58 58 52.7 & --52.7 & 690 & $<$0.8 &  5.7 & $<$1.E--04 &   \\
322.158+0.636 & 15 18 34.61 & --56 38 24.9 & --50 & 178 & $<$1.6 & $<$9.9 & $<$7.E--04 & $<$2.E--04 \\
323.740--0.263 & 15 31 45.44 & --56 30 50.1 & --53.1 & 3000 & $<$0.9 &  12.5 & $<$2.E--05 &   \\
326.475+0.703 & 15 43 16.73 & --54 07 14.6 & --50 & 110 & $<$1.5 & $<$1.8 & $<$1.E--03 & $<$6.E--05 \\
327.120+0.511 & 15 47 32.73 & --53 52 38.4 & --103.1 & 90 & $<$0.9 &  9.2 & $<$8.E--04 &   \\
327.291--0.578 & 15 53 07.79 & --54 37 06.8 & --50 & 2.6 & $<$1.5 &   & $<$5.E--02 &   \\
328.237--0.547 & 15 57 58.32 & --53 59 22.9 & --50 & 360 & $<$1.6 & $<$0.9 & $<$4.E--04 & $<$1.E--05 \\
328.254--0.532 & 15 57 59.78 & --53 58 00.8 & --50 & 440 & $<$1.5 & $<$2.6 & $<$3.E--04 & $<$2.E--05 \\
328.808+0.633 & 15 55 48.45 & --52 43 06.5 & --53.5 & 240 & $<$1.0 &  5.5 & $<$3.E--04 &   \\
329.029--0.205 & 16 00 31.81 & --53 12 49.7 & --50 & 138 & $<$1.4 & $<$1.2 & $<$9.E--04 & $<$3.E--05 \\
335.726+0.191 & 16 29 27.36 & --48 17 53.2 & --50 & 78 & $<$1.5 & $<$1.4 & $<$2.E--03 & $<$7.E--05 \\
335.789+0.174 & 16 29 47.33 & --48 15 51.7 & --50 & 118 & $<$1.3 & $<$0.8 & $<$9.E--04 & $<$3.E--05 \\
336.018--0.827 & 16 35 09.26 & --48 46 47.4 & --54.1 & 400 & $<$0.8 &  6 & $<$2.E--04 &   \\
337.705--0.053 & 16 38 29.62 & --47 00 35.5 & --50 & 120 & $<$1.4 & $<$2.6 & $<$1.E--03 & $<$8.E--05 \\
339.622--0.121 & 16 46 05.99 & --45 36 43.3 & --50 & 95 & $<$1.4 & $<$1.4 & $<$1.E--03 & $<$6.E--05 \\
339.884--1.259 & 16 52 04.67 & --46 08 34.1 & --40 & 1650 & $<$0.6 &  56 & $<$3.E--05 &   \\
340.054--0.244 & 16 48 13.88 & --45 21 43.6 & --54 & 42 & $<$0.8 &  2.9 & $<$2.E--03 &   \\
340.785--0.096 & 16 50 14.86 & --44 42 26.4 & --103.9 & 144 & $<$0.7 &  6.1 & $<$4.E--04 &   \\
341.218--0.212 & 16 52 17.83 & --44 26 52.2 & --50 & 137 & $<$1.4 & $<$0.6 & $<$8.E--04 & $<$2.E--05 \\
344.227--0.569 & 17 04 07.78 & --42 18 39.5 & --15 & 68 & $<$1.6 & $<$3.7 & $<$2.E--03 & $<$2.E--04 \\
345.003--0.223 & 17 05 10.89 & --41 29 06.2 & --24.4 & 240 & $<$0.6 &  3.5 & $<$2.E--04 &   \\
345.010+1.792 & 16 56 47.57 & --40 14 25.7 & --10 & 410 & $<$0.5 &  82 & $<$9.E--05 &   \\
345.504+0.348 & 17 04 22.88 & --40 44 22.3 & --24.2 & 172 & $<$0.8 &  2.3 & $<$4.E--04 &   \\
348.550--0.979 & 17 19 20.39 & --39 03 51.8 & --15 & 74 & $<$2.0 & $<$0.9 & $<$2.E--03 & $<$5.E--05 \\
348.703--1.043 & 17 20 04.05 & --38 58 30.8 & --24.7 & 60 & $<$0.9 &  7.6 & $<$1.E--03 &   \\
348.727--1.037 & 17 20 06.55 & --38 57 09.1 & --15 & 90 & $<$2.0 & $<$3.4 & $<$2.E--03 & $<$1.E--04 \\
351.417+0.645 & 17 20 53.37 & --35 47 01.2 & --10 & 2600 &  32.8 &  14.8 &   &   \\
351.445+0.660 & 17 20 54.60 & --35 45 08.6 & --15 & 100 & $<$1.5 & $<$11.7 & $<$1.E--03 & $<$5.E--04 \\
351.775--0.536 & 17 26 42.57 & --36 09 17.6 & --24 & 230 & $<$0.7 & $<$12.4 & $<$2.E--04 & $<$2.E--04 \\
352.630--1.067 & 17 31 13.90 & --35 44 08.8 & --15 & 160 & $<$2.0 & $<$1.8 & $<$1.E--03 & $<$4.E--05 \\
353.410--0.360 & 17 30 26.18 & --34 41 45.5 & --24.3 & 87 & $<$0.8 &  5.5 & $<$8.E--04 &   \\
354.615+0.472 & 17 30 17.09 & --33 13 55.0 & --15 & 151 & $<$2.0 & $<$1.6 & $<$1.E--03 & $<$4.E--05 \\
0.645--0.042 & 17 47 18.67 & --28 24 24.8 & 60 & 69 & $<$1.5 &   & $<$2.E--03 &   \\
9.621+0.196 & 18 06 14.67 & --20 31 32.3 & --10 & 5090 & $<$0.6 &  22 & $<$9.E--06 &   \\
10.473+0.027 & 18 08 38.20 & --19 51 50.1 & 70 & 61 & $<$0.6 & $<$10.4 & $<$8.E--04 & $<$7.E--04 \\
12.909--0.260 & 18 14 39.52 & --17 51 59.9 & 40 & 317 & $<$0.8 &  5.5 & $<$2.E--04 &   \\
23.010--0.411 & 18 34 40.26 & --09 00 38.2 & 70 & 405 & $<$0.9 &  5.2 & $<$2.E--04 &   \\
23.440--0.182 & 18 34 39.18 & --08 31 24.3 & 100 & 77 & $<$0.9 &  4.4 & $<$1.E--03 &   \\
35.201--1.736 & 19 01 45.54 & 01 13 35.1 & 30.5 & 560 & $<$0.9 &  24 & $<$1.E--04 &   \\
\end{tabular}}                          

\end{table*}


\section{RESULTS}

The 23.1-GHz methanol maser spectrum detected in 351.417+0.645 (NGC~6334F) is shown in Fig~\ref{fig:ngc_2}.  The maser spectrum is well fitted by 3 Gaussian components (see Table~\ref{tab:gfit}).  The 32.8~Jy main component is at -10.39~\kms;  this may be compared with the detection in 1987 by Menten \& Batrla of a 51.6~Jy signal at -10.36~\kms.  Variability is a key characteristic of maser emissions, and here a decrease of one third in absolute intensity is apparent over the 14 years between the two observations.  A second peak near -11~\kms\ is clearly visible in our spectrum, with no significant change in the relative intensity over the 6 days of our observations.  In the Menten \& Batrla observation only one component was identified, but a second component of the same relative intensity could be present within the noise.      

No 23.1-GHz methanol emission was detected from any of the remaining 49 sources.  Upper limits ($3\sigma$, Jy) on the flux density $S$ are reported in Table~\ref{tab:sources}, together with the centre of the velocity range searched for each source.  In Table~\ref{tab:sources} the 6.6-GHz maser peak intensity and the 107.0-GHz maser peak intensity (20 sources) or upper limit (28 sources) are also tabulated (Caswell \etal\ 2000).  In sources which display thermal emission at 107.0~GHz the thermal peak intensity is taken as the upper limit on maser emission.  Note that when masers are detected at both 107.0 and 6.6~GHz, the peaks do not always coincide in velocity, although the 107.0-GHz maser always falls within the velocity range of the 6.6-GHz maser.  In the following analysis we make no quantitative comparison between the detected 6.6- and 107.0-GHz maser peaks to avoid assumptions about the coincidence and relative size of these emissions.  

Without knowing the sizes of individual 6.6-GHz maser sources we cannot estimate their absolute brightness temperatures.  But where masers are not detected, the absence of emission at 23.1 or 107.0~GHz sets an upper limit on the relative brightness temperatures in the 6.6-GHz maser source, which may be compared with the results of maser modelling.    Columns headed $R$ evaluate the following ratios.
\[
R(23.1/6.6) = [S(23.1)/S(6.6)](6668/23121)^{2}
\]
\[
R(107.0/6.6) = [S(107.0)/S(6.6)](6668/107013)^{2}
\]
These represent the upper limits to the brightness temperature ratios $T_{b}(23.1)/T_{b}(6.6)$ and $T_{b}(107.0)/T_{b}(6.6)$ respectively, for 23.1- or 107.0-GHz emission with the same angular extent as the peak 6.6-GHz maser component. 

For the 23.1-GHz maser detected in 351.417+0.645 we find $R(23.1/6.6)=10^{-3}$, while for the remaining 49 6.6-GHz maser sources the upper limits on $R(23.1/6.6)$ range from $<9\times10^{-6}$ to $<5\times10^{-2}$.  For the 28 6.6-GHz methanol maser sources in Table~\ref{tab:sources} for which we have upper limits at both 23.1 and 107.0~GHz, we find that limits  on $R(23.1/6.6)$ range between $<10^{-4}$ and $<6\times10^{-3}$, while limits on $R(107.0/6.6)$ range between $<10^{-5}$ and $<7\times10^{-4}$.  In Section~4 we use these two constraints to investigate conditions in the 6.6-GHz maser sources.


\section{MASER MODELLING}

Previously we have used maser intensity ratios for several detected lines in an individual source to set constraints on the physical conditions in that source.  Here we adopt a complementary approach which looks at the range of plausible conditions for the set of sources surveyed, with the aim of setting constraints on the conditions in the majority of class~II methanol maser sources.  Rather than trying to explain the phenomenon of the bright 23.1-GHz masers in just 2 sources, we instead investigate what the absence of these masers tells us about the more typical maser environment.


\subsection{The maser model}

We have used the maser pumping model of Sobolev \& Deguchi (1994) to map out the prevalence of the 6.6-, 23.1- and 107.0-GHz methanol masers as a function of the model parameters representing physical conditions.  In this model the methanol molecules are excited by infrared photons emanating from a layer of warm dust, assumed to be heated by emission from the nearby newborn star, or perhaps by a shock.  Inverted populations result from the radiative and collisional cascade back to the ground state in the (generally) cooler gas.  Masers develop when there is sufficient column density of methanol in the gas phase, with a large degree of velocity coherence along the line of sight.  They may be augmented by a beamed geometry, or the amplification of background continuum radiation.

The maser pumping is governed by infrared radiation from warm dust at temperature \Td, which promotes the methanol molecules to their torsionally excited states.  We assume a dust filling factor of 0.5 and opacity $(\nu/10^{4})^2$, where $\nu$ is the frequency in GHz.  \Td\ above 100~K is sufficient to activate the torsional pumping.

The masers develop in gas of temperature \Tk\ and volume density \nH, where the methanol molecules undergo collisions with molecular hydrogen.  Most of the masers are strongest in cool gas (i.e. \Tk\ much less than \Td), although masers persist even when $\Tk > \Td$ if there is sufficient optical depth.  Increasing gas density eventually quenches the masers through thermalization.

Collisional excitation rates are based on the model of Peng \& Whiteoak (1993), which uses propensity rules derived by Lees \& Haque (1974) from a small number of laboratory measurements.  Calculated rate coefficients for rotational excitation of methanol by helium have recently become available (Pottage, Flower \& Davis 2002).  These authors found their calculated rates to be consistent with the Lees \& Haque propensity rules.  Unfortunately the Pottage \etal\ rates extend only up to J=9, so do not include many of the significant processes for levels involved in the 23.1-GHz $9_{2}-10_{1}\ \rm{A}^{+}$ transition which is the subject of this paper.  Therefore we have retained the propensity rule treatment of collisions.  While accurately calculated rates for the full range of J would provide more confidence in the quantitative aspects of the maser model, the pumping is dominated by radiative processes.  For example, we have compared model calculations involving nonselective collisions with those based on the experimental propensity rules (Sobolev \etal\ 1997a, 1997b, Cragg \etal\ 2001).  There is no change to the masers in the low density limit, and as the density is increased the same masers are evident in both cases.  Quantitative details such as the maser peak intensities and the densities at which they become thermally quenched are influenced by the collision model.  Rates of collisional excitation to the first torsionally excited state have recently been calculated for the first time by Pottage, Flower \& Davis (2004).  The cross-sections for these transitions are typically two orders of magnitude smaller than for transitions within the torsional ground state, and exhibit no clear propensity for particular changes in J or K, suggesting that collisional excitation between torsional states can be safely neglected.   

Radiative transfer is treated in the large velocity gradient (LVG) approximation, modified by a beaming factor $\beam=10$, defined as the ratio of optical depths parallel and perpendicular to the line of sight.  This represents the assumed elongation of the maser region towards the observer.  Beaming allows large optical depths to develop in the maser transitions along the line of sight, while the pumping radiation is incident from the perpendicular direction, so the pumping transitions can remain optically thin.  In this treatment the maser amplification and hence the brightness is governed by \beam\scdm, representing the column density of methanol along the line of sight divided by the line width of the emission.  The parameter \scdm\ is referred to as the specific column density of methanol.  

The two symmetry species of methanol, A and E, correspond to different nuclear spin states, and are not interconverted by the radiative and collisional processes considered here.  Therefore they are modelled independently.  The A and E species are likely to be of equal abundance, [A]=[E], unless the molecules form at very low temperatures, in which case there is an excess of the A species.  The parameter \scdm\ applies to a single symmetry species.

As an example, consider a model with parameters $\beam=10$, $\nH=10^{7}$~\cc, and $\scdm=10^{12}$~\ccs.  This could be realized by assorted different combinations of maser path length $L$ and methanol abundance $X$.  For instance, if [A]=[E], $\DV=0.5$~\kms, and the length of the amplifying path is $L=10^{17}$~cm, then the total column density of methanol along the line of sight is $10^{18}\ \rm cm^{-2}$ and the fractional abundance of methanol relative to hydrogen is $X=10^{-6}$.  

To model the observed brightness temperatures, \scdm\ must be large enough to give appreciable optical depth, but note from the above example that very large values of \scdm\ are only realistically achievable at very high gas densities \nH.  When [A]=[E] the relationship between the various quantities may be written as follows.
\[
{\rm log}(\scdm) = {\rm log}(\nH) + {\rm log}(XL/(2\beam\DV))
\]
High resolution observations such as those by Caswell (1997) and Phillips \etal\ (1998) show that the maximum extent of the maser spot distribution at a single maser site is approximately $\sim 6000$~AU ($10^{17}$~cm or 30~milliparsec).  If we assume that we do not lie along a preferred line of sight, then the linear size of the maser cluster gives a reasonable estimate of the size of the masing gas cloud, and hence the maximum path length $L$.  Methanol fractional abundances as large as $4\times 10^{-6}$ have been obtained towards quasi-thermal cores in SgrB2 (Mehringer \& Menten 1997), and we set $X<10^{-5}$ as an upper limit.  Observations by Caswell \etal\ (1995a, 1995b) and Moscadelli \etal\ (2003) demonstrate that methanol maser line widths are unlikely to be less than 0.15~\kms.  Combining these limits with the assumed beaming factor $\beam=10$ in our models gives
\[
{\rm log}(\scdm) < {\rm log}(\nH) + 6.5
\]
for plausible models.  Parameter combinations which don't satisfy this inequality represent models with excessive maser path lengths or methanol abundance relative to hydrogen, and are described as unrealistic conditions in the following sections.

Enhanced methanol abundances are believed to result from the evaporation of icy grain mantles in the vicinity of a young star (Hartquist \etal\ 1995).  The results of Sandford \& Allamandola (1993) suggest that methanol may start to come off the grains at an appreciable rate at temperatures above 80~K, while the phase transition which occurs at 120~K will selectively eject methanol into the gas phase.  In the model calculations many of the masers become strongest when the gas temperature is much lower than this, and we examine temperatures as low as 25~K.  However, it is unlikely that the methanol-rich gas required for maser amplification is present at such low temperatures.

Previous calculations with this model allowed the masers to amplify continuum radiation from an underlying \HII\ region, although this is not necessary for the production of the masers.  Less than half the class~II methanol maser sources have associated ultra-compact \HII\ regions, and it is suggested that the majority of masers accompany the very early stages of high-mass star formation, before the development of a detectable uc\HII\ continuum (e.g. Phillips \etal\ 1998, Walsh \etal\ 1998).  Here we omit the \HII\ background continuum in the interests of reducing the number of adjustable model parameters.  Note that both exceptional sources in which strong 23.1-GHz masers have been detected, W3(OH) and NGC~6334F, contain well developed and bright uc\HII\ regions.  In this paper we do not attempt detailed modelling of these sources, which must of course include such radiation.

In the new calculations reported here we examine the effects of varying four of the model parameters: dust temperature \Td, gas temperature \Tk, gas volume density \nH, and methanol specific column density \scdm.  The conclusions drawn are subject to the deficiencies in the model treatment of radiative transfer, dust opacity and collisional excitation, as outlined in Appendix B of Sutton \etal\ (2001).


\begin{figure}
\centerline{\epsfxsize=8.5cm\epsfbox{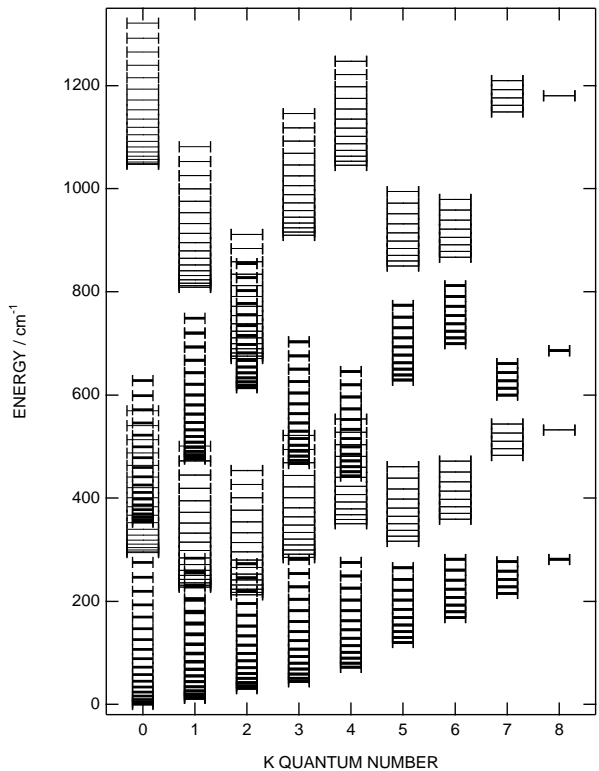}}
\caption{Energy levels used in maser model calculations for the A symmetry species of methanol.  Levels of ground and first 3 torsionally excited states are included, up to J= 18.  For clarity, $\vt=0$ and 2  are shown as heavier and narrower ladders compared with $\vt=1$ and 3.}
\label{fig:levels_A}
\end{figure}


\subsection{Methanol energy levels}

The gross features of the distribution of population among the various methanol energy levels are governed by the two model temperatures:  the dust temperature controlling how molecules are transferred between torsional states, while the gas temperature affects the spread of population among the rotational levels within each torsional state.  Of course it is the fine features of the population distribution which produce the masers, and it is important to avoid distorting these by excessive truncation of the energy level set.

Sobolev \& Deguchi (1994) found that pumping to the second torsionally excited state ($\vt=2$) was necessary to account for the extreme brightness of the 12.1-GHz maser, and their calculations included methanol energy levels for $\vt=0,1,2$.  Allowed transitions between different torsional states have the selection rules $\Delta\rm{J}=0,\pm1$, $\Delta\rm{K}=\pm1$.  The torsionally excited states hold only a small fraction of the total population, since molecules which are elevated from the ground state by absorption of infrared photons then decay rapidly back to the ground state in a few steps.  Nevertheless it is necessary to include a comparable number of rotational levels in each accessible torsional state since all populated ground state levels may participate in the infrared absorption.

We have recently looked further into the question of how many energy levels should be included in the modelling.  As expected, the higher energy levels participate at higher temperatures.  The results of this work will be reported in detail elsewhere.  We found the 23.1-GHz maser to be somewhat sensitive to the $\vt=3$ levels, and so have included them for the first time in the calculations reported here.  Trial calculations with larger data sets suggest that the energy levels included in this work give adequate results for dust temperatures $\Td<300$~K and gas temperatures $\Tk<200$~K.  Line strengths for transitions involving $\vt=3$ levels are not accurately known, and we have used approximate values based on the product of symmetric rotor line strengths with the torsional overlap.

Fig~\ref{fig:levels_A} shows the energy levels included in our calculations for the A symmetry species of methanol based on the data of Mekhtiev, Godfrey \& Hougen (1999).  Levels of the ground and first 3 torsionally excited states are included; these are separated by approximately 200 $\rm cm^{-1}$.  Within the ground torsional state levels up to 280 $\rm cm^{-1}$ are included; these are sorted by K quantum number and displayed as ladders of increasing J, with levels up to J=18 included.  The  corresponding levels in the torsionally excited states are also included.  The irregular position of the energy ladders in the excited states is a result of the large torsion-rotation interaction in this molecule.  The A symmetry species has close pairing of levels, which are labelled by a $\pm$ symmetry label, while the E species levels are instead labelled by a signed K quantum number.  In total our calculations include 760 levels of A species methanol, and were repeated for 748 levels of E species methanol.


\begin{figure}
\centerline{\epsfxsize=7cm\epsfbox{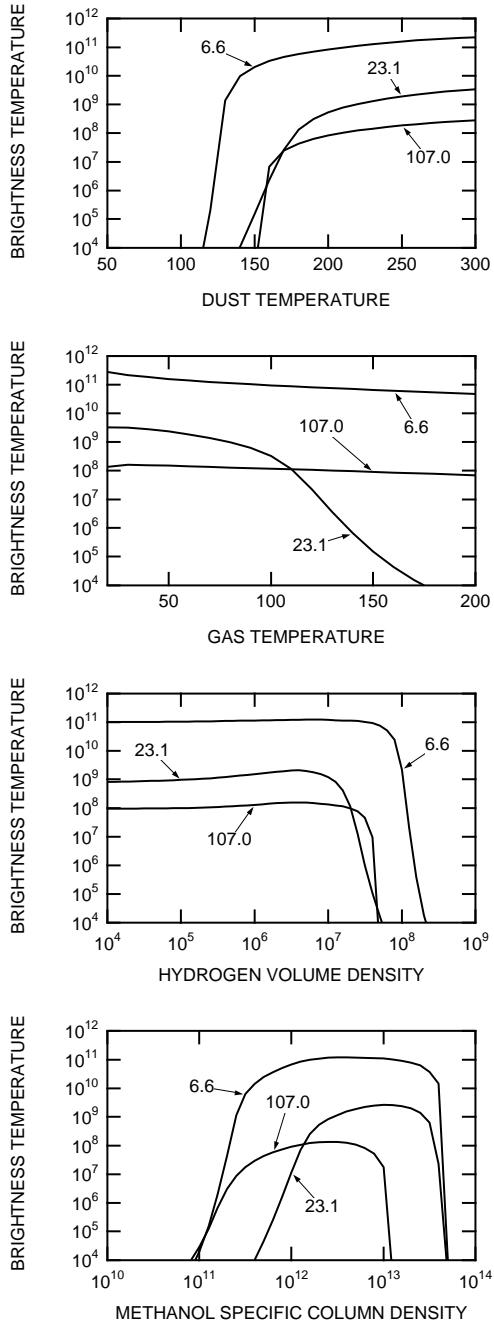}}
\caption{Change of brightness temperature as model parameters are varied one-at-a-time, for masers at 6.6, 23.1 and 107.0~GHz (as labelled).  In each panel one parameter is varied while the rest are fixed at the following values: dust temperature $\Td=225$~K, gas temperature $\Tk=75$~K, gas volume density $\nH=10^{7}$~\cc, and methanol specific column density $\scdm=10^{12.5}$~\ccs.}
\label{fig:vary_1D}
\end{figure}


\subsection{Variation of a single parameter}

As a starting point we consider conditions $\Td=225$~K, $\Tk=75$~K, $\nH=10^{7}$~\cc, $\scdm=10^{12.5}$\ccs, with the remaining model parameters fixed as described in Section~4.1.  Under these conditions the model predicts strong masers in all three of the 6.6-, 23.1- and 107.0-GHz lines, so such conditions are certainly NOT typical of actual methanol maser sources.  The behaviour of the three masers as each of the four parameters is varied individually is shown in Fig~\ref{fig:vary_1D}.  The masers switch on when the pumping is sufficiently strong at dust temperatures $\Td>100$~K, and switch off due to thermalization at gas densities $\nH>10^{8}$~\cc.  The 23.1-GHz maser is strongest in cool gas, switching off when $\Tk>150$~K under the chosen conditions, while the 6.6- and 107.0-GHz masers are less influenced by gas temperature.  The masers are very bright for methanol specific column densities in the range $10^{11}<\scdm<10^{14}$~\ccs, with the 23.1-GHz maser requiring a somewhat greater column density of methanol to become active.  The maser brightness temperatures saturate under favourable conditions, but can rise and fall very rapidly as conditions change.  The peak brightness is greatest for the 6.6-GHz maser, and under the conditions illustrated is greater for the 23.1-GHz maser than the 107.0-GHz maser.

It is apparent from these plots that the model can readily account for the simultaneous presence of masers at 6.6, 23.1 and 107.0~GHz, as has been observed in 3 sources to date.  There are also a range of parameter combinations for which the 6.6-GHz maser is active, with or without the 107.0-GHz maser, but no maser is apparent at 23.1~GHz.  Although the 23.1-GHz masers are rare, they feature in the model under quite a wide range of conditions, and so the absence of these masers is consistent with only a restricted range of parameters.  In the following sections we map out these parameter combinations to delimit the physical conditions in the majority of maser sources surveyed.  Here we do not attempt to delineate conditions in sources where the 23.1- and/or 107.0-GHz masers were detected.


\subsection{Variation of two parameters}

\begin{figure}
\centerline{\epsfxsize=7cm\epsfbox{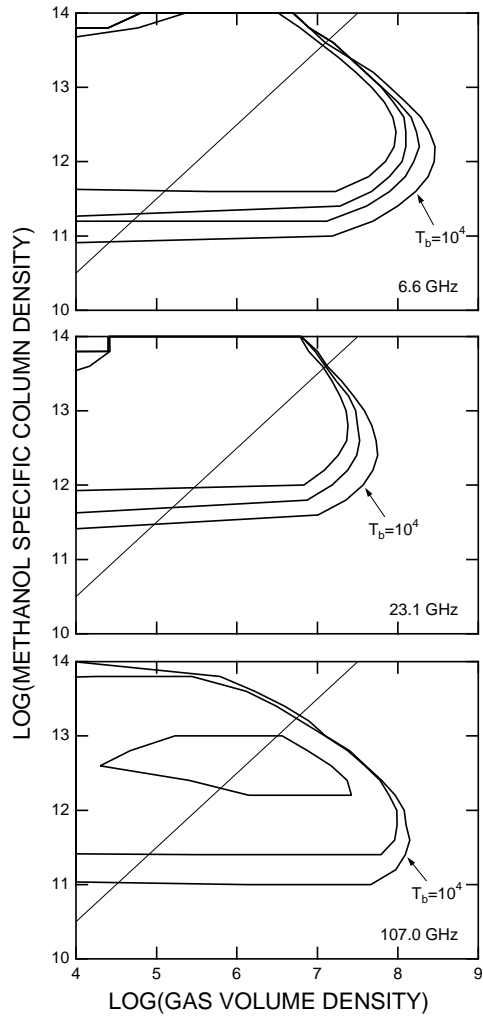}}
\caption{Contours of maser brightness temperature as \nH\ and \scdm\ are varied, for $\Tk=75$~K and $\Td=225$~K.  Contour values are $\Tb=10^{4}$, $10^{6}$, $10^{8}$ and $10^{10}$~K.  The line log(\scdm)=log(\nH)+6.5 distinguishes plausible model conditions (lower right) from those requiring unrealistically large maser path lengths or methanol abundance (upper left). }
\label{fig:contours}
\end{figure}

Fig~\ref{fig:contours} shows contours of maser brightness temperature for the 6.6-, 23.1- and 107.0-GHz lines as a function of two model parameters, the gas density and methanol specific column density.  Here the dust and gas temperatures are fixed at $\Td=225$~K and $\Tk=75$~K respectively.  The closeness of the contours denotes the rapid rise and fall of maser brightness already seen in Fig~\ref{fig:vary_1D}.  It is evident that all three masers are prevalent in the regime of  low gas density and high methanol specific column density, while the 23.1-GHz maser is active over a slightly narrower range of parameters than the others.

In Fig~\ref{fig:contours} the upper left portion of each plot represents models with log(\scdm) $>$ log(\nH)+6.5 which require unrealistically large maser path lengths or methanol abundance relative to hydrogen, as discussed in Section 4.1.  Since 6.6-GHz brightness temperatures are likely to exceed $10^{8}$~K, the top panel of Fig~\ref{fig:contours} shows that plausible maser models which account for the 6.6-GHz emission span several orders of magnitude in \nH\ and \scdm, the range being a function of \Tk\ and \Td.


\begin{figure}
\centerline{\epsfxsize=7cm\epsfbox{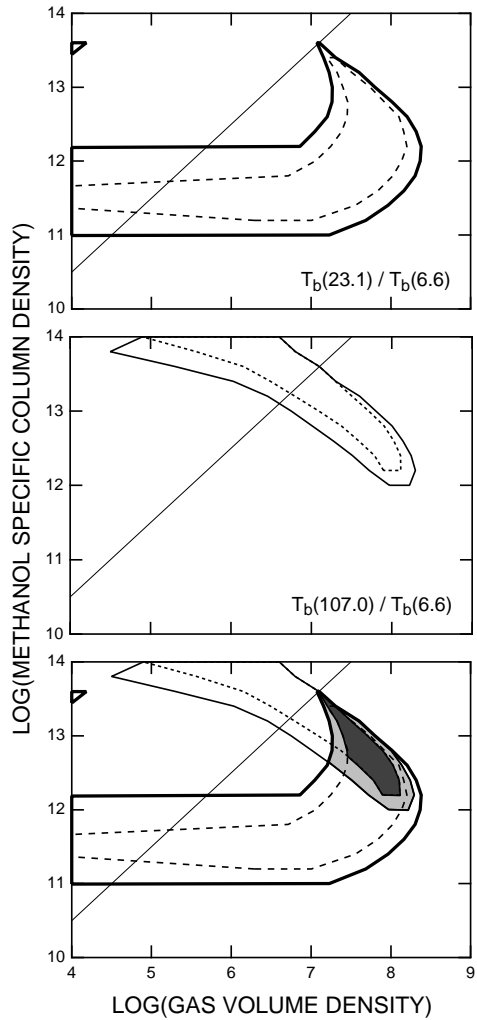}}
\caption{Ratios of maser brightness temperature as \nH\ and \scdm\ are varied, for $\Tk=75$~K and $\Td=225$~K.  First and second panels show contours of relative brightness temperature of 23.1- and 107.0-GHz masers respectively, in relation to 6.6-GHz maser brightness, while third panel shows the overlap.   Solid and dotted contours represent weakest and strongest upper limits derived from survey data, for sources in which neither maser is detected.  The line log(\scdm)=log(\nH)+6.5 distinguishes plausible model conditions (lower right) from those requiring unrealistically large maser path lengths or methanol abundance (upper left).}
\label{fig:ratios}
\end{figure}


The range of model conditions compatible with the 23.1-GHz upper limits from our survey can be mapped out by comparing the ratios of brightness temperatures $\Tb(23.1)/\Tb(6.6)$ from the model with the upper limits on $R(23.1/6.6)$ as evaluated in Table~\ref{tab:sources}.  Many combinations of model conditions can be ruled out, because as well as producing maser action at 6.6~GHz, they inevitably also produce masers at 23.1~GHz, in contrast to the observations.  The same can be said of masers at 107.0~GHz for the majority of the 6.6-GHz maser sources.  

The top panel of Fig~\ref{fig:ratios} shows contours of the ratio of brightness temperatures $\Tb(23.1)/\Tb(6.6)$ as gas density and methanol column density are varied, again for $\Td=225$~K and $\Tk=75$~K.  The contour values represent the weakest ($<6\times10^{-3}$) and strongest ($<10^{-4}$) upper limits of $R(23.1/6.6)$ from the 28 sources in our survey in which no 23.1-GHz methanol emission was detected, and for which upper limits at 107.0~GHz are also available.  The methanol specific column density is restricted to values in the range $10^{11}<\scdm<10^{12}$~\ccs\ for hydrogen density $\nH<10^{6.5}$~\cc, but may extend to greater values up to $\scdm=10^{14}$~\ccs\ for some values of hydrogen density in the range $10^{6.5}<\nH<10^{8}$~\cc.  

The middle panel of Fig~\ref{fig:ratios} shows contours of $\Tb(107.0)/\Tb(6.6)$, over the same range of parameters.  The contour values represent the weakest ($<7\times10^{-4}$) and strongest ($<10^{-5}$) upper limits of $R(107.0/6.6)$ for the same 28 sources, as observed by Caswell \etal\ (2000).  The methanol specific column density is restricted to values $\scdm>10^{12}$~\ccs, while the required value of hydrogen density decreases as the methanol specific column density increases.

Note that the range of conditions defined by each of these ratios fall within the regime in which the 6.6-GHz transition is a maser, with the outer boundary of the regions defined in Fig ~\ref{fig:ratios} lying very close to the $\Tb(6.6)=10^{5}$~K contour (Fig~\ref{fig:contours}).  Thus the observed relative intensities at 6.6, 23.1 and 107.0~GHz exclude both quasi-thermal conditions, and conditions where the 23.1- and/or 107.0-GHz masers become very bright.

In the bottom panel of Fig~\ref{fig:ratios} the shaded area represents model conditions which satisfy both requirements, that is both the 23.1- and 107.0-GHz emission is low enough (in relation to the 6.6-GHz emission) to meet the survey criteria.  Models with parameters in the lightly shaded area meet the weakest limits set by the survey data, while the darker shaded area defines model conditions satisfying the most stringent observational limits.  Thus within the shaded region are model conditions which will permit upper limits to be satisfied for all 28 sources in which the 6.6-GHz masers lack counterparts at higher frequencies (although not all points in the shaded region will satisfy the limits obtained in all sources).  In this case all models which satisfy the survey criteria also fall in the zone of plausible source size and methanol abundance.  A further restriction on acceptable model parameters may be obtained by arguing that the 6.6-GHz brightness temperature should exceed $10^{8}$~K (rather than $10^{5}$~K as achieved in the shaded region), but the closeness of the contours in the relevant region (Fig~\ref{fig:contours}) means that the effect is slight.


\begin{figure*}
\centerline{\epsfxsize=18cm\epsfbox{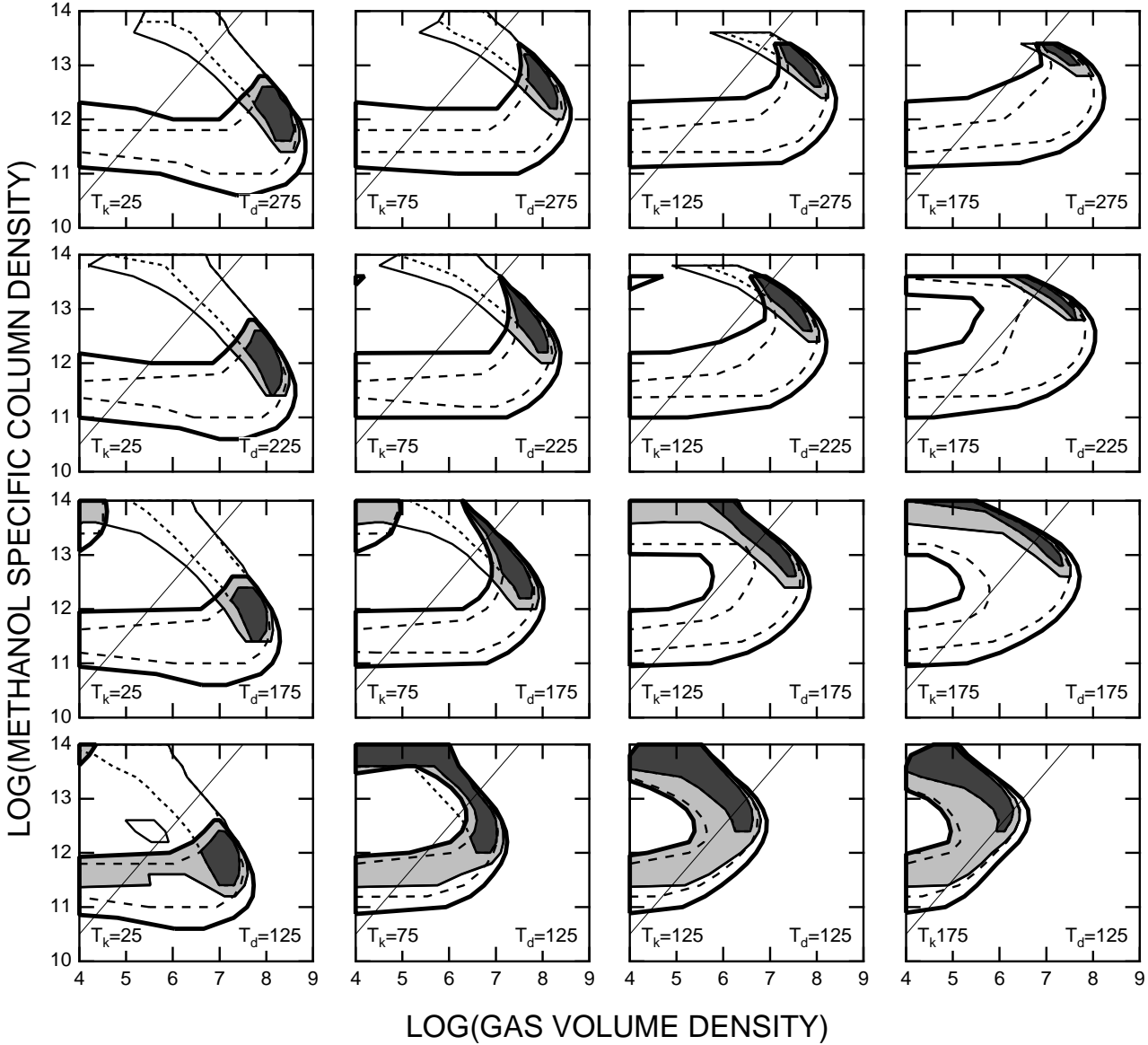}}
\caption{Ratios of maser brightness temperature as four model parameters are varied.  Heavy and dashed contours delimit the parameter combinations compatible with the strongest and weakest upper limits at 23.1~GHz, while light and dotted contours do the same for 107.0~GHz.  The shaded areas represent conditions satisfying upper limits for both transitions, for sources with masers detected at 6.6~GHz but not at 23.1 and 107.0~GHz.  In each panel the line log(\scdm)=log(\nH)+6.5 distinguishes plausible model conditions (lower right) from those requiring unrealistically large maser path lengths or methanol abundance (upper left).}
\label{fig:ratios_4D}
\end{figure*}


\subsection{Variation of four parameters}

It is apparent that the 23.1- and 107.0-GHz limits can provide complementary information, and together can set severe restrictions on the conditions in typical 6.6-GHz maser sources, where there are no observable 23.1- and 107.0-GHz maser counterparts.  For a more comprehensive investigation we have repeated the analysis for dust temperatures $\Td=125, 175, 225, 275$~K and gas temperatures $\Tk=25, 75, 125, 175$~K, giving 16 combinations in all.  Fig~\ref{fig:ratios_4D} shows the resulting plots, with the shaded regions again representing conditions which can satisfy the observed upper limits at both 23.1 and 107.0~GHz.  Each plot gives results for a different combination of gas and dust temperatures, with gas temperatures increasing from left to right and dust temperatures increasing from bottom to top.  Within each plot the gas volume density varies along the horizontal axis while the methanol specific column density varies along the vertical axis.  The brightness temperature of the 6.6-GHz maser ranges between $4\times 10^{4}$ and $4\times 10^{11}$~K for models in the shaded regions.  Again, unrealistic model parameter combinations appear in the top left zone of each panel.  Some models which meet the observational limits fall within the unrealistic zone.

The 23.1-GHz methanol maser is readily excited at low gas temperatures.  The same is true of the stronger and more widespread 6.6- and 107.0-GHz masers.  All three also extend to gas temperature 175~K if the other model conditions are right.  For the 23.1-GHz maser, this requires more extreme combinations of high dust temperature, low gas density and/or high methanol column density than for the other two transitions.  The absence of masers at 23.1~GHz is most restrictive on the other model parameters at low gas temperatures, while the absence of masers at 107.0~GHz is most restrictive at high gas temperatures.  Both limits impose stronger restrictions at higher dust temperatures.

At dust temperatures $175 \leq \Td \leq 275$~K there are rather limited combinations of \nH\ and \scdm\ which satisfy both the observational limits and the reasonable source size and methanol abundance criteria.  These are confined to gas density $10^{6.5} \leq \nH \leq 10^{8.5}$~\cc\ and methanol specific column densitiy $10^{11.5} \leq \scdm \leq 10^{13.5}$~\ccs, with \scdm\ values at the low end of this range required at $\Tk=25$~K and \scdm\ values at the high end of the range at $\Tk=175$~K.

At more moderate dust temperatures - above the minimum required to activate the masers but below 175~K - the range of \nH\ and \scdm\ combinations compatible with the observed intensity limits and reasonable model criteria shifts to lower values: $10^{5}  \leq \nH \leq 10^{7.5}$~\cc\ and $10^{11.5} \leq \scdm \leq 10^{13}$~\ccs.  In the panel with $\Tk=175$~K and $\Td=125$~K there is very little scope for models which meet the reasonable source size and methanol abundance criteria, so that gas temperatures exceeding the dust temperature are unlikely to represent actual maser conditions.  


\section{DISCUSSION}

Our survey of 50 southern class~II methanol maser sources detected maser emission at 23.1~GHz in only one source.  It is apparent that methanol masers at 23.1~GHz are rare, or if they exist, are generally very much weaker than those found in W3(OH) (10~Jy) and NGC~6334F (32.8~Jy).   The NGC~7538 previous maser detection at 0.5~Jy is comparable with our upper limits.  Note that the 23.1~GHz maser is present towards NGC~6334F and W3(OH), two sources known to exhibit a large number of methanol maser transitions, which suggests that conditions which produce the 23.1~GHz maser may also be suitable for producing other rare maser transitions.  The third source with class~II methanol maser emission detected at many frequencies is 345.01+1.79, in which we obtained a $3\sigma$ upper limit of 0.5~Jy.

In earlier work we fitted the methanol maser model to observations available at the time in 345.01+1.79 and NGC~6334F (Cragg \etal\ 2001) and W3(OH) (Sutton \etal\ 2001).   All three exhibit methanol maser emission at 107.0~GHz.  These models included continuum radiation from a background \HII\ region, so differ in detail from those presented here.  Nevertheless, a comparison with Fig~\ref{fig:ratios_4D} shows that the physical conditions represented by models which offered approximate fits to these observations are distinct from those derived for the majority of 6.6-GHz maser sources, in which maser emission is not detected at either 23.1 or 107.0~GHz.  

In particular, our previous modelling of several methanol masers in W3(OH) found gas temperature exceeding 110~K (Sutton \etal\ 2001).  This analysis was based on the site of peak maser emissions at 86.6, 86.9 and 107.0~GHz, which BIMA observations located near the northern edge of the uc\HII\ region.  In contrast, the VLA observations of Menten \etal\ (1998a) found the strongest 23.1-GHz maser spot approx 1.5\arcsec\ away near the southern edge of the uc\HII\ region, with only weak 23.1-GHz emission corresponding to the northern site.  Thus these two sites exhibit assorted methanol masers with very different intensity ratios, which presumably reflects different local physical conditions.  Our survey demonstrates that the conditions required for strong 23.1-GHz emissison are uncommon in methanol maser sources, W3(OH) and NGC~6334F being the exceptions.  The model suggests that either the combination of gas temperature, dust temperature, gas density and methanol column density is extreme (e.g., the gas at these locations has cooled, perhaps to 75 K) or some other pecularities unaccounted by our simple model come into effect.  Further modelling (which must include the effects of the background uc\HII\ continuum radiation in these sources) may help define the physical conditions at these sites, but is best undertaken after high resolution observations have clarified the associations between the various masers - for example, the locations of the strong masers at 19.9~GHz.  In the current paper we are attempting to interpret the emissions (or lack thereof) in the more typical 6.6-GHz methanol maser sources without associated masers at 23.1 or 107.0~GHz.

Of the several hundred class~II maser sites detected at 6.6~GHz, more than 100 also show maser emission at 12.1~GHz, while masers at 107.0~GHz have been detected in only 25, and masers in the remaining class~II methanol lines are each known in only a few sources.  Our survey results are consistent with the picture of a typical class~II maser source showing significant maser emission only at 6.6 and (possibly) 12.1~GHz.  The sources which also have masers at 107.0~GHz form an especially interesting set, some also showing maser emission in additional lines.  Multitransition analysis offers the possibility of discriminating between the conditions in these sources which give rise to different combinations of observed maser lines.  The upper limits at 23.1~GHz from our survey for 19 of the 25 known 107.0-GHz maser sources will provide further constraints on such analyses.

For the class~II methanol maser sources with emission at 6.6 and/or 12.1~GHz the maser models have defined general conditions necessary for maser action: infrared radiation from dust at temperature above 100~K, gas density not exceeding $10^{8.5}$~\cc, and enhanced methanol abundance relative to hydrogen $>10^{-8}$.  But because these two masers are prevalent in the models over a wide range of parameters, the observations at 6.6 and 12.1~GHz alone are not sufficient to narrowly define conditions in the typical class~II maser sources.  The presence or absence of masers at other frequencies can usefully subdivide this quite broad parameter space.  The maser models have been successful in identifying new maser candidates, and offer the potential for modelling conditions in individual 107.0-GHz maser sources once surveys at several frequencies are complete.

We have shown in the preceding section that the upper limits on maser emission in the higher frequency lines can help delineate the range of possible conditions in the large number of sources in which maser action is confined to 6.6 and 12.1~GHz.  This follows from the prevalence of such high frequency masers in the models.  The combination of upper limits at 23.1 and 107.0~GHz with the observed 6.6-GHz maser peak intensity has proved fruitful.  The current study suggests that the typical 6.6~GHz methanol maser sources form under conditions which are either near the high gas density limit for methanol masers, or are only just above the dust temperature threshold for maser pumping.  It also demonstrates that the observed 6.6-GHz masers need special conditions of one sort or another - if the dust temperature does not exceed the gas temperature, then rather high levels of methanol abundance $X > 10^{-6}$ and long maser paths $\sim 10^{17}$~cm are required.

Most of the sources observed in our survey also exhibit methanol maser emission in the 12.1-GHz line.  This is in accord with the model, in which the 6.6- and 12.1-GHz transitions become strong masers simultaneously over a wide range of model conditions.  The 6.6-GHz maser is nearly always stronger than the 12.1-GHz maser in both observations and modelling.  However, the 12.1-GHz maser does extend to higher gas densities in some models, which may account for the cases in which the 12.1-GHz maser is observed to be the stronger of the two.  Our sample contains 26 sources with masers detected at 12.1~GHz (Caswell \etal\ 1995b) for which we have upper limits at 23.1 and 107.0~GHz.  Note that the 23.1- and 107.0-GHz upper limits apply separately to each of the 6.6- and 12.1-GHz maser peaks, even when these are different components at different velocities.  Repeating the analysis of the last section for the brightness temperature ratios $\Tb(23.1)/\Tb(12.1)$ and $\Tb(107.0)/\Tb(12.1)$ produces a rather similar result, except that the constraints on model parameters are less strict than those obtained in Fig~\ref{fig:ratios_4D} from $\Tb(23.1)/\Tb(6.6)$ and $\Tb(23.1)/\Tb(6.6)$.  Specifically the dark shaded region (tightest limit) for the 6.6-GHz ratios always lies within the equivalent region for the 12.1-GHz ratios.  The light shaded region (loosest limit) for the 6.6-GHz ratios also lies within the equivalent region for the 12.1-GHz ratios - except at $\Td=125$~K where the 6.6-GHz ratios permit slightly lower values of methanol specific column density \scdm.  This independent treatment of the 12.1-GHz maser observations is consistent with our 6.6-GHz analysis, but does not add much in the way of further constraints.  More information can potentially be obtained by combining the 6.6- and 12.2-GHz maser observations in cases where they are likely to be cospatial, but we have avoided that assumption in this paper.


\section{CONCLUSION}

A new search for 23.1-GHz methanol masers in 50 southern star formation regions has confirmed the previously detected maser emission in NGC~6334F, but found no new masers at this frequency.  The scarcity of 23.1~GHz masers in the target sources, which all exhibit class~II methanol maser emission at 6.6~GHz, provides new constraints for models of the maser pumping.  Further constraints apply in sources with upper limits to maser emission at 107.0~GHz.  Combining data for the two transitions sets limits on the physical conditions in sources representative of the majority of methanol maser sites in the vicinity of newborn high-mass stars, subject to the uncertainties of the applied model.  Conditions are likely to be near the limits of 6.6~GHz maser activity, with gas density $10^{6.5}-10^{8.5}$~\cc\ near the upper threshold, and/or dust temperature $100-150$~K near the lower threshold.


\section*{Acknowledgements}

The Australia Telescope is funded by the Commonwealth of Australia for operation of a National Facility managed by CSIRO.  Financial support for this work was provided by the Australian Research Council and the Victorian Partnership for Advanced Computing.  AMS thanks the Australian Research Council for an IREX Award which funded travel to Australia, and acknowledges financial support by INTAS and the Ministry of Industry, Science and Technology of the Russian Federation (Contract No. 40.022.1.1.1102).



\end{document}